\documentclass[seceq]{ptptex}

\usepackage{graphicx}

\notypesetlogo                       

\title{
Parquet theory for correlated thermodynamic Greens functions%
}

\author{
Peter \textsc{Kleinert}%
}

\inst{
Paul-Drude Institut f\"{u}r Festk\"{o}rperelektronik, Hausvogteiplatz 5-7, 10117 Berlin, Germany.
}

\abst{
We address the problem of finding self-consistent parquet equations for two-body correlated Greens functions that arise out of a cumulant expansion. The general theory as developed for non-relativistic electron systems reproduces the conventional two-body parquet diagrammatic technique for the vertex functions. As an application, the crossing-invariant linearized basic parquet equations are treated for the correlated two-particle Greens function.
}

\PTPindex{062}  

\begin{document}

\maketitle

\section{Introduction}
Parquet equations are used to self-consistently sum up ring and ladder diagrams that are needed for the microscopic description of numerous many-body systems including the Coulomb gas, liquid helium, and nuclear matter. These equations were originally derived by Landau et al. \cite{Landau} to study the nonperturbative domain in quantum electrodynamics and in the theory of two-particle scattering to construct an extension of the Bethe-Salpeter equation \cite{Diatlov} that accounts for the crossing symmetry, which is a consequence of the Pauli principle. This symmetry is easily described, but extremely difficult to implement in practical applications. For a two-particle Greens function (GFs) that refers to Fermions, this fundamental symmetry is expressed by
\begin{equation}
G(12,1^{\prime}2^{\prime})=-G(21,1^{\prime}2^{\prime})=-G(12,2^{\prime}1^{\prime})=G(21,2^{\prime}1^{\prime}),
\end{equation}
where the numbers collect space, imaginary time, and spin indices (e.g., 1 stands for the space $\boldsymbol{r}_1$, time $t_1$, and spin $s_1$ coordinates). It is important that this simple symmetry couples different scattering channels to each other by means of partial irreducible vertex functions. This coupling is accounted for by a closed set of nonlinear self-consistent equations for the two-particle irreducible vertices, which are called parquet equations. The theory is applicable not only for small but also for large values of the coupling constants and allows a simultaneous treatment of long-range as well as short-range correlations \cite{Jackson_1982,Bickers_1991,He_1993,Jackson_1994}.

Applications of the parquet approach start from an initial set of irreducible diagrams, which cannot be broken into two parts by cutting a pair of electron lines to generate all related reducible diagrams of the respective $n$-body GF. The full parquet equations, which are based on all irreducible diagrams, are exact, in accordance with the crossing symmetry, and conserving with respect to the particle number, the energy, the momentum, and the angular momentum. Unfortunately, this internal consistency of the full parquet theory breaks down in any approximation \cite{Jackson_1987}. Since the rigorous solution of the full equations is not feasible for even the simplest models, it is necessary to treat simplified, manageable variants of the originally exact full theory. Several diagrammatic approximations beyond perturbation theory have been proposed in the literature and successfully applied to the treatment of many-body problems \cite{Weiner,Jackson_85,Zhele,Janis_2005,Toschi,Janis_2008}. A further systematic extension of all these techniques is possible by a calculational scheme, which is based on a rearrangement of the whole diagrammatic collection. Such a transformation of the entire hierarchy of equations for the many-body GFs is carried out by using an alternative generating functional. Of particular interest in statistical physics is the cumulant expansion, which provides a natural framework for the definition of an $n$-particle approximation as being the set of equations for correlation functions that terminates, when cumulants of the order $n+1$ are discarded. Applied to the many-particle theory, the cumulant expansion generates equations of motion for correlated thermodynamic GFs that is nothing but a rearrangement of the Kadanoff-Baym hierarchy \cite{Martin_Schwinger}. It is the advantage of this cumulant approach that it provides a firm basis for a systematic self-consistent $n$-particle approach by cutting the chain of equations at the $(n+1)$-particle level. This rearranged diagrammatic expansion has to satisfy the crossing symmetry, which is a fundamental property of any many-particle system. Unfortunately, the resulting set of equations violates the crossing symmetry. This observation motivates our work to propose a system of crossing-invariant parquet equations for correlated thermodynamic GFs that are particularly useful for the construction of $n$-particle approximations.

\section{Correlated Green functions}
Let us develop the many-body theory for an electron gas in a solid that is described by the Hamiltonian
\begin{eqnarray}
&&H=\sum\limits_{s}\int {\rm d}\boldsymbol{r}\psi_s^{\dag}(\boldsymbol{r},t)\left[-\frac{\hbar^2}{2m}\Delta_{\boldsymbol{r}}+V(\boldsymbol{r}) \right]
\nonumber\\
&&+\frac{1}{2}\sum\limits_{s,s^{\prime}}\int {\rm d}\boldsymbol{r}{\rm d}\boldsymbol{r}^{\prime}
\psi_s^{\dag}(\boldsymbol{r},t)\psi_{s^{\prime}}^{\dag}(\boldsymbol{r}^{\prime},t)v(\boldsymbol{r}-\boldsymbol{r}^{\prime})
\psi_{s^{\prime}}(\boldsymbol{r}^{\prime},t)\psi_s(\boldsymbol{r},t),
\label{Hamil}
\end{eqnarray}
with $\psi_s^{\dag}(\boldsymbol{r},t)$ [$\psi_s(\boldsymbol{r},t)$] denoting the creation [annihilation] operator for a Fermion with spin $s$ that occupies the lattice site $\boldsymbol{r}$ at time $t$. The carriers with an effective mass $m$ move in the single-particle potential $V(\boldsymbol{r})$ and interact with each other via the Coulomb potential $v(\boldsymbol{r}-\boldsymbol{r^{\prime}})$. The equal-time anticommutator relations of the field operators have the form
\begin{subequations}
\begin{eqnarray}
&&\left[\psi_{s}(\boldsymbol{r},t),\psi_{s^{\prime}}(\boldsymbol{r}^{\prime},t) \right]_{+}
=\left[\psi_{s}^{\dag}(\boldsymbol{r},t),\psi_{s^{\prime}}^{\dag}(\boldsymbol{r}^{\prime},t) \right]_{+}=0,\\
&&\left[\psi_{s}(\boldsymbol{r},t),\psi_{s^{\prime}}^{\dag}(\boldsymbol{r}^{\prime},t) \right]_{+}=\delta_{s,s^{\prime}}
\delta(\boldsymbol{r}-\boldsymbol{r}^{\prime}),
\label{Commutator}
\end{eqnarray}
while the time dependence of the system is governed by the von Neumann equation
\end{subequations}
\begin{equation}
-i\hbar\frac{\partial}{\partial t}\psi_{s}(\boldsymbol{r},t)=\left[H,\psi_{s}(\boldsymbol{r},t) \right]_{-},
\end{equation}
which is used to derive equations of motion for the GFs. A straightforward approach is based on the generating functional
\begin{eqnarray}
G[\lambda,\eta]=&&1+\sum\limits_{n=1}^{\infty}\frac{1}{(n!)^2}\int {\rm d}1\dots {\rm d}n{\rm d}1^{\prime}\dots {\rm d}n^{\prime}\lambda(n)\dots\lambda(1)\nonumber\\
&&\times G(1\dots n,1^{\prime}\dots n^{\prime})\eta(1^{\prime})\dots\eta(n^{\prime}),
\label{DefG}
\end{eqnarray}
in which both $\eta(j)$ and $\lambda(j)$ denote anticommutating functions. The infinite chain of coupled equations for the $n$-particle GFs $G(1\dots n,1^{\prime}\dots n^{\prime})$ is generated from a functional differential equation for $G[\lambda,\eta]$
\begin{eqnarray}
&&\biggl\{
\frac{\hbar^2}{2m}\Delta_{\boldsymbol{r}_1}-V(\boldsymbol{r}_1)+i\hbar\frac{\partial}{\partial t_1}
\biggl\}\frac{\delta}{\delta \lambda(1)}G[\lambda,\eta]\label{funcab}\\
&&=\eta(1)G[\lambda,\eta]-i\hbar\int {\rm d}\overline{1}V(1-\overline{1})\frac{\delta}{\delta\eta(\overline{1}^{+})}
\frac{\delta}{\delta\lambda (\overline{1})}\frac{\delta}{\delta\lambda (1)}G[\lambda,\eta],\nonumber
\end{eqnarray}
with $V(1-2)=v(\boldsymbol{r}_1-\boldsymbol{r}_2)\delta(t_1-t_2)$. An alternative hierarchy of equations is derived for correlated $n$-body GFs $G_c(1\dots n,1^{\prime}\dots n^{\prime})$ by introducing its generating functional via the equation
\begin{equation}
G[\lambda,\eta]=\exp\left(G_c[\lambda,\eta] \right).
\label{eq1}
\end{equation}
Exploiting Eq.~(\ref{funcab}), the exact results for the one-particle and two-particle correlated GFs are easily derived \cite{Martin_Schwinger}. The first equations of this infinite chain have the form
\begin{equation}
G(1,1^{\prime})=G_0(1,1^{\prime})-i\hbar\int{\rm d}\overline{1}{\rm d}\overline{2}
G_0(1,\overline{1})V(\overline{1}-\overline{2})G_c(\overline{2}\overline{1},\overline{2}^{+}1^{\prime}),
\label{glG}
\end{equation}
\begin{subequations}
\begin{eqnarray}
&&G_c(12,1^{\prime}2^{\prime})=i\hbar\int{\rm d}\overline{1}{\rm d}\overline{2} V(\overline{1}-\overline{2})
\biggl\{-G_c(\overline{2}\overline{1}2,\overline{2}^{+}1^{\prime}2^{\prime})\label{gla}\\
&&+\left[G_0(1,\overline{1})G(2,\overline{2})G(\overline{1},1^{\prime})G(\overline{2},2^{\prime})
-G_0(1,\overline{1})G(2,\overline{2})G(\overline{1},2^{\prime})G(\overline{2},1^{\prime}) \right]\label{glb}\\
&&+\left[G_0(1,\overline{1})G(2,\overline{2})G_c(\overline{1}\overline{2},1^{\prime}2^{\prime}) \right]\label{glc}\\
&&+\left[G_0(1,\overline{1})G(\overline{2},1^{\prime})G_c(\overline{1}2,\overline{2}^{+}2^{\prime})
+G_0(1,\overline{1})G(\overline{2},2^{\prime})G_c(\overline{1}2,1^{\prime}\overline{2}^{+}) \right]\label{gld}\\
&&-\left[G_0(1,\overline{1})G(\overline{1},1^{\prime})G_c(\overline{2}2,\overline{2}^{+}2^{\prime})
+G_0(1,\overline{1})G(\overline{1},2^{\prime})G_c(\overline{2}2,1^{\prime}\overline{2}^{+}) \right]\biggl\},\label{gle}
\end{eqnarray}
\end{subequations}
where the appearance of the three-particle GF in Eq.~(\ref{gla}) should be noted. It is a special feature of this set of equations for the cumulants that the Hartree-Fock GF $G_0$ appears in each equation of the hierarchy exactly once in each term. $G_0$ itself is determined by a integro-differential equation
\begin{eqnarray}
&&\left[\frac{\hbar^2}{2m}\Delta_{\boldsymbol{r}_1}-V(\boldsymbol{r}_1)+i\hbar\int{\rm d}\overline{1} V(1-\overline{1})G(\overline{1},\overline{1}^{+})
+i\hbar\frac{\partial}{\partial t} \right]G_0(1,1^{\prime})\nonumber\\
&&-i\hbar\int{\rm d}\overline{1} V(1-\overline{1})G(1,\overline{1}^{+})G_0(\overline{1},1^{\prime})=\delta(1-1^{\prime}),
\label{glG0}
\end{eqnarray}
which includes the full one-particle GF $G$ in a self-consistent manner. Taking into account the relationship
\begin{equation}
G(12,1^{\prime}2^{\prime})=G_c(12,1^{\prime}2^{\prime})+G(1,1^{\prime})G(2,2^{\prime})-
G(1,2^{\prime})G(2,1^{\prime})
\end{equation}
between the standard and correlated two-particle GF, we conclude that the effective one-particle approach as given by the Hartree-Fock approximation results, when the correlated two-particle propagator $G_c(12,1^{\prime}2^{\prime})$ is neglected. Accordingly, an optimized pair approximation is established, when the three-particle GF in Eq.~(\ref{gla}) is removed. The resulting closed equation for the correlated two-particle GF differs from the related standard result by the appearance of the GF $G_0$. Indeed, by replacing $G_0$ by $G$ in Eqs.~(\ref{gla}) to (\ref{gle}), we recover the linearized basic parquet equation, in which only the lowest-order irreducible vertex function is taken into account. Besides this formal correspondence, the crossing symmetry is basically violated in Eq.~(2.8). This serious defect of the cumulant expansion has to be removed before this powerful approach can be exploited. What is needed is a crossing invariant theory for correlated many-particle GFs.
%

\section{Parquet equations}
The full imaginary-time parquet equations self-consistently generate all reducible Feynman diagrams for the two-body GF in terms of the bare interaction, the fully dressed one-particle GF, and an initial set of irreducible diagrams by maintaining the complete crossing symmetry. The crossing operation itself transfers reducible vertex diagrams of a given scattering channel into irreducible vertices of complementary channels so that a coupled, non-linear set of self-consistent integral equations arises for the irreducible two-particle vertices. To apply this calculational procedure to the hierarchy of correlated GFs, we chose a mixed representation, in which not only vertex functions appear, but also partial two-particle GFs. Accordingly, the expression for the total correlated four-point GF $G_c$ decomposes into a sum of different scattering contributions
\begin{equation}
G_c=g_0+g_{ee}+g_d+g_c.
\label{pa1}
\end{equation}
The collection of all irreducible vertex diagrams $\Gamma_{\rm irr}$ leads to the component $g_0$ given by
\begin{equation}
g_0(12,1^{\prime}2^{\prime})=\frac{1}{2}\int{\rm d}\overline{1}\dots{\rm d}\overline{4}
\Lambda_0(12,\overline{1}\overline{2})
\Gamma_{\rm irr}(\overline{1}\overline{2},\overline{3}\overline{4})
G(\overline{3},1^{\prime})G(\overline{4},2^{\prime}),
\label{pa2}
\end{equation}
where $\Lambda_0$ denotes the symmetrized product of one-particle GFs
\begin{equation}
\Lambda_0(12,1^{\prime}2^{\prime})=G_0(1,1^{\prime})G(2,2^{\prime})+G(1,1^{\prime})G_0(2,2^{\prime}).
\end{equation}
Particle-particle ladder diagrams are summed in $g_{ee}$, which satisfies the self-consistent integral equation
\begin{equation}
g_{ee}(12,1^{\prime}2^{\prime})=\frac{1}{4}\int{\rm d}\overline{1}\dots{\rm d}\overline{4}
\Lambda_0(12,\overline{1}\overline{2})
\Gamma_{ee}(\overline{1}\overline{2},\overline{3}\overline{4})
G_c(\overline{3}\overline{4},1^{\prime}2^{\prime}).
\label{pa3}
\end{equation}
The construction of the particle-hole hierarchy is more involved. To account for the crossing symmetry, we make use of the decomposition
\begin{equation}
g_{d,c}=\frac{1}{2}\left(g_{d,c}^{(1)}+g_{d,c}^{(2)} \right),
\label{pa0}
\end{equation}
where the components satisfy the respective integral equations
\begin{equation}
g_d^{(1)}(12,1^{\prime}2^{\prime})=-\int{\rm d}\overline{1}\dots{\rm d}\overline{4}
G_0(1,\overline{1})G(\overline{3},1^{\prime})\Gamma_d(\overline{1}\overline{2},\overline{3}\overline{4})
G_c(\overline{4}2,\overline{2}2^{\prime}),
\label{pa4}
\end{equation}
\begin{equation}
g_c^{(1)}(12,1^{\prime}2^{\prime})=\int{\rm d}\overline{1}\dots{\rm d}\overline{4}
G_0(1,\overline{1})G(\overline{4},2^{\prime})\Gamma_c(\overline{1}\overline{2},\overline{3}\overline{4})
G_c(\overline{3}2,1^{\prime}\overline{2}).
\label{pa5}
\end{equation}
The quantities $g_{d,c}^{(2)}$ are obtained from the right-hand side of these equations by interchanging the numbers 1 with 2 as well as 1$^{\prime}$ with 2$^{\prime}$. The vertex functions, which enter Eqs.~(\ref{pa3}), (\ref{pa4}), and (\ref{pa5}), are given by
\begin{equation}
\Gamma_{ee}=\Gamma_{\rm irr}+\gamma_d+\gamma_c,\quad
\Gamma_d=\Gamma_{\rm irr}+\gamma_c+\gamma_{ee},\quad
\Gamma_c=\Gamma_{\rm irr}+\gamma_d+\gamma_{ee},
\label{defgam}
\end{equation}
where the partial vertices $\gamma_{\alpha}$ (with $\alpha=c$, $d$, or $ee$) generate the GFs $g_{\alpha}$ via the relationship
\begin{equation}
g_{\alpha}(12,1^{\prime}2^{\prime})=\int{\rm d}\overline{1}\dots{\rm d}\overline{4}
G(1,\overline{1})G(2,\overline{2})\gamma_{\alpha}(\overline{1}\overline{2},\overline{3}\overline{4})
G(\overline{3},1^{\prime})G(\overline{4},2^{\prime}).
\label{pa6}
\end{equation}
The set of Eqs.~(\ref{pa1}) to (\ref{pa6}) together with Eqs.~(\ref{glG}) and (\ref{glG0}) establish the self-consistent parquet approach for the calculation of the correlated one- and two-particle GFs. An iteration of these equations can start from an effective expression for the Hartree-Fock GF $G_0$ and an initial set of irreducible vertex diagrams $\Gamma_{\rm irr}$ that is used to calculate the first approximation for $G_c$ from Eqs.~(\ref{pa1}) and (\ref{pa2}). To proceed further, also the partial vertex functions $\gamma_d$, $\gamma_c$, and $\gamma_{ee}$ have to be determined from the integral Eq.~(\ref{pa6}). The solution of these equations is facilitated by a Fourier transformation with respect to the spatial coordinates and by a transformation to Matsubara frequencies $z_{\nu}$ according to
\begin{eqnarray}
G_c(12,1^{\prime}2^{\prime})&&=\left(\frac{i}{\hbar\beta} \right)^3\sum\limits_{\nu,\nu^{\prime},n}
G_c(x_1x_2,x_{1^{\prime}}x_{2^{\prime}}|z_{\nu}z_{\nu^{\prime}}\omega_n)\\
&&\times \exp\left[-iz_{\nu}t_1-i(\omega_n-z_{\nu})t_2+iz_{\nu^{\prime}}t_{1^{\prime}}
+i(\omega_n-z_{\nu^{\prime}})t_{2^{\prime}}\right],\nonumber
\end{eqnarray}
where $\beta=1/(k_BT)$ denotes the temperature parameter and $z_{\nu}=i\pi\nu/(\hbar\beta)+\mu/\hbar$, $\omega_n=i\pi n/(\hbar\beta)+2\mu/\hbar$, with $\mu$ being the chemical potential. The result is an algebraic relationship
\begin{eqnarray}
&&g_{s_1s_2,s_3s_4}(\boldsymbol{k}\boldsymbol{k}^{\prime}\boldsymbol{q}|z_{\nu}z_{\nu^{\prime}}\omega_n)=
G(\boldsymbol{k},z_{\nu})G(\boldsymbol{q}-\boldsymbol{k},\omega_n-z_{\nu})\nonumber\\
&&\times \gamma_{s_1s_2,s_3s_4}(\boldsymbol{k}\boldsymbol{k}^{\prime}\boldsymbol{q}|z_{\nu}z_{\nu^{\prime}}\omega_n)
G(\boldsymbol{k}^{\prime},z_{\nu^{\prime}})G(\boldsymbol{q}-\boldsymbol{k}^{\prime},\omega_n-z_{\nu^{\prime}}),
\end{eqnarray}
which is easily inverted. Furthermore, the $\gamma_c$ diagrams are obtained from the ones for $\gamma_d$ by exploiting the antisymmetry of the vertex function under the exchange of the electron (or hole) variables. This specific crossing symmetry is expressed by
\begin{equation}
\gamma_{d}(12,1^{\prime}2^{\prime})=-\gamma_{c}(12,2^{\prime}1^{\prime}).
\end{equation}
The set of parquet Eqs.~(\ref{pa1}) to (\ref{pa6}) for the correlated two-body GF is the main result of this paper. The solution of these equations together with Eqs.~(\ref{glG}) and (\ref{glG0}) is a formidable task that is generally not possible without further approximations. Therefore, we discuss two limiting cases, which have received particular attention.

First, let us derive the conventional parquet equations for GFs that have been studied in a number of review articles \cite{Jackson_1982,Bickers_1991,He_1993,Jackson_1994}. As already mentioned, the main peculiarity in the theory of correlated GFs is the appearance of $G_0$. As the linearized basic parquet equations are obtained from Eq.~(2.8) by the replacement $G_0\rightarrow G$, one expects a similar unfolding for the set of Eqs.~(\ref{pa1}) to (\ref{pa6}). Indeed, replacing $G_0$ by $G$ in these parquet equations, we recover closed equations for the total vertex function $\Gamma$, which is defined by Eq.~(\ref{pa6}), with $g_{\alpha}$ and $\gamma_{\alpha}$ being replaced by $G_c$ and $\Gamma$, respectively. Obvious manipulations lead to the final equations
\begin{equation}
\Gamma=\Gamma_{\rm irr}+\gamma_{ee}+\gamma_d+\gamma_c,
\end{equation}
\begin{equation}
\gamma_d(12,1^{\prime}2^{\prime})=-\int{\rm d}\overline{1}\dots{\rm d}\overline{4}
\Gamma_d(1\overline{1},1^{\prime}\overline{2})G(\overline{2},\overline{3})G(\overline{4},\overline{1})
\Gamma(\overline{3}2,\overline{4}2^{\prime}),
\end{equation}
\begin{equation}
\gamma_c(12,1^{\prime}2^{\prime})=\int{\rm d}\overline{1}\dots{\rm d}\overline{4}
\Gamma_c(1\overline{1},\overline{2}2^{\prime})G(\overline{2},\overline{3})G(\overline{4},\overline{1})
\Gamma(\overline{3}2,1^{\prime}\overline{4}),
\end{equation}
\begin{equation}
\gamma_{ee}(12,1^{\prime}2^{\prime})=\frac{1}{2}\int{\rm d}\overline{1}\dots{\rm d}\overline{4}
\Gamma_{ee}(12,\overline{1}\overline{2})G(\overline{1},\overline{3})G(\overline{2},\overline{4})
\Gamma(\overline{3}\overline{4},1^{\prime}2^{\prime}),
\end{equation}
which agree with the conventional parquet equations for the vertex function \cite{Weiner}. Again, the vertices $\Gamma_c$, $\Gamma_d$, and $\Gamma_{ee}$ of different scattering channels appear, which are defined by Eq.~(\ref{defgam}). The exact correspondence between the parquet equations for the two different sets of many-body GFs (namely $G$ and $G_c$) provides strong evidence for our crossing invariant result in Eqs.~(\ref{pa1}) to (\ref{pa6}).

\begin{figure}
\centerline{\includegraphics[width=10cm,height=10cm]
{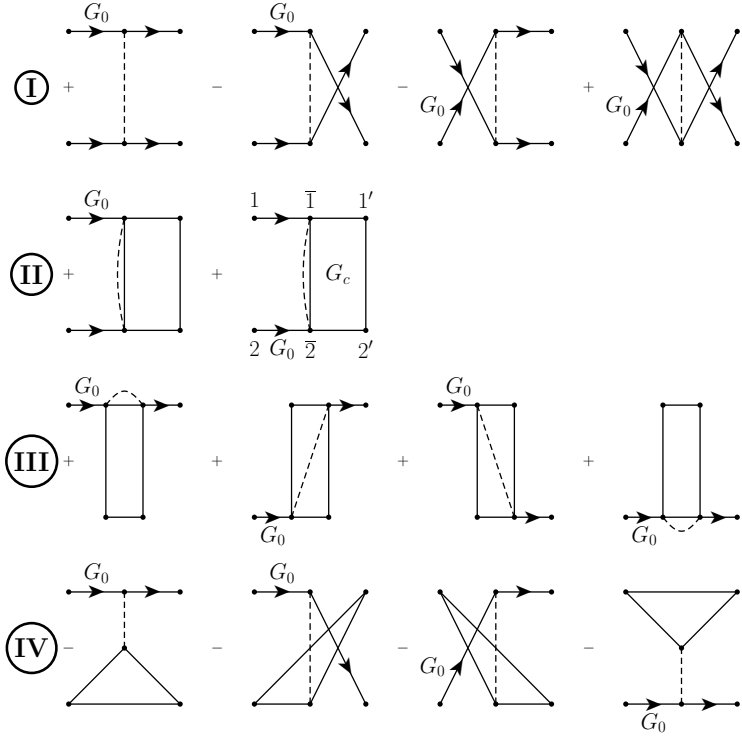}}
\caption{Feynman diagrams for the correlated two-particle GF $G_c(12,1^{\prime}2^{\prime})$ arranged in four rows I to IV, which collect subgroups that satisfy the crossing symmetry. The bare potential $V$ is represented by a dashed line. One-particle propagators $G$ and $G_0$ are shown by oriented solid lines.}
\label{fig:1}
\end{figure}

For practical applications, a linearization of the full parquet equations is sometimes desirable. A further simplification consists in a restriction to the lowest-order irreducible vertex function, which is given by
\begin{equation}
\Gamma_0(12,1^{\prime}2^{\prime})=i\hbar V(1-2)\left[
\delta(1-1^{\prime})\delta(2-2^{\prime})-\delta(1-2^{\prime})\delta(2-1^{\prime}) \right].
\end{equation}
Adopting both approximations, the parquet Eqs.~(\ref{pa1}) to (\ref{pa6}) lead to an equation for the correlated two-particle GF, which is diagrammatically represented in Fig. 1. The collection of all diagrams in the rows I to IV satisfies the crossing symmetry and has, therefore, the potential to describe distinct physical entities. Particle-particle and particle-hole ladder diagrams are generated by the channels II and (III, IV), respectively. A mixing of spin states appears in the electron-hole channel IV, which encompasses components that originate from the vertex functions $\gamma_d$ and $\gamma_c$. The equation for $G_c$ as represented by the diagrams in Fig. 1 is nothing but a crossing invariant extension of Eq.~(2.8), which was straightforwardly derived from the generating functional $G_c[\lambda,\eta]$.

To further discuss the physical significance of our result, let us treat two examples in more detail. The set of diagrams shown in Fig. 1 by row I fulfill the crossing symmetry. All contributions are composed from one-particle GFs that are combined in the GW approximation for the self-energy. In the Fourier space, we have
\begin{equation}
\Sigma(\boldsymbol{k},z_{\nu})=G_0^{-1}(\boldsymbol{k},z_{\nu})-G^{-1}(\boldsymbol{k},z_{\nu})
=\frac{1}{\beta}\sum\limits_{\boldsymbol{q},n}W(\boldsymbol{k},\boldsymbol{q}|\omega_n)
G(\boldsymbol{q}-\boldsymbol{k},\omega_n-z_{\nu}),
\label{gw1}
\end{equation}
\begin{eqnarray}
W(\boldsymbol{k},\boldsymbol{q}|\omega_n)=&&\frac{1}{\beta}\sum\limits_{\boldsymbol{k}_1,\nu_1}
\left[v(\boldsymbol{k}-\boldsymbol{k}_1)^2+v(\boldsymbol{k}+\boldsymbol{k}_1-\boldsymbol{q})^2
-v(\boldsymbol{k}-\boldsymbol{k}_1)v(\boldsymbol{k}+\boldsymbol{k}_1-\boldsymbol{q}) \right]\nonumber\\
&&\times G_0(\boldsymbol{k}_1,z_{\nu_1})G(\boldsymbol{q}-\boldsymbol{k}_1,\omega_n-z_{\nu_1}).
\label{gw2}
\end{eqnarray}
This result can be viewed as a natural two-particle extension of the Hartree-Fock approximation ($G=G_0$), which trivially satisfies the crossing symmetry. The full self-consistent GW approximation in Eqs.~(\ref{gw1}) and (\ref{gw2}) has the special feature that the kernel $W$ depends on the squared coupling constant as well as on the Hartree-Fock GF $G_0$.

More interesting is the crossing symmetric Cooper channel II, which collects all relevant diagrams of the BCS theory of superconductivity. As the $V$ dependent parts of $\gamma_{ee}$ are never large \cite{Weiner}, we restrict ourselves to the homogeneous integral equation, which is expressed by
\begin{eqnarray}
&&G_c(12,1^{\prime}2^{\prime})= i\hbar\int{\rm d}\overline{1}{\rm d}\overline{2} V(\overline{1}-\overline{2})
\label{Cooper}\\
&&\times\frac{1}{2}\left[G_0(1,\overline{1})G(2,\overline{2})+G(1,\overline{1})G_0(2,\overline{2}) \right]
G_c(\overline{1}\overline{2},1^{\prime}2^{\prime})\nonumber.
\end{eqnarray}
An analytic solution of this equation together with Eq.~(\ref{glG}) is possible by exploiting the symmetry relation
\begin{equation}
G_c(x_1x_2,x_{1^{\prime}}x_{2^{\prime}}|z_{\nu}z_{\nu^{\prime}}\omega_n)=
-G_c^{*}(x_{1^{\prime}}x_{2^{\prime}},x_1x_2|z_{\nu^{\prime}}^{*}z_{\nu}^{*}\omega_n^{*}),
\end{equation}
(with $x_j=\boldsymbol{r}_j,\,s_j$) and by introducing the gap function $\Delta$ according to
\begin{equation}
\Delta(x_1x_2,x_{1^{\prime}}x_{2^{\prime}}|\omega_n)=v(\boldsymbol{r}_1-\boldsymbol{r}_2)
v(\boldsymbol{r}_{1^{\prime}}-\boldsymbol{r}_{2^{\prime}})
\frac{1}{\beta^2}\sum\limits_{\nu,\nu^{\prime}}G_c(x_1x_2,x_{1^{\prime}}x_{2^{\prime}}|z_{\nu}z_{\nu^{\prime}}\omega_n).
\end{equation}
Applying a spatial Fourier transformation, the coupled Eqs.~(\ref{glG}) and (\ref{Cooper}) are expressed by
\begin{eqnarray}
&&G(\boldsymbol{k},z_{\nu})=G_0(\boldsymbol{k},z_{\nu})-G_0(\boldsymbol{k},z_{\nu})
\frac{i}{\hbar\beta}\sum\limits_{\boldsymbol{q},n}e^{\varepsilon\omega_n}\Delta(\boldsymbol{k}\boldsymbol{k}^{\prime}\boldsymbol{q}|\omega_n)\\
&&\times\frac{1}{2}\left[G_0(\boldsymbol{k},z_{\nu})G(\boldsymbol{q}-\boldsymbol{k},\omega_n-z_{\nu})
+G(\boldsymbol{k},z_{\nu})G_0(\boldsymbol{q}-\boldsymbol{k},\omega_n-z_{\nu}) \right],\nonumber
\end{eqnarray}
\begin{eqnarray}
&&\Delta(\boldsymbol{k}\boldsymbol{k}^{\prime}\boldsymbol{q}|\omega_n)=\sum\limits_{\boldsymbol{k}_1}v(\boldsymbol{k}-\boldsymbol{k}_1)
\Delta(\boldsymbol{k}_1\boldsymbol{k}^{\prime}\boldsymbol{q}|\omega_n)\\
&&\times\biggl\{-\frac{1}{2\beta}\left[G_0(\boldsymbol{k}_1,z_{\nu})G(\boldsymbol{q}-\boldsymbol{k}_1,\omega_n-z_{\nu})
+G(\boldsymbol{k}_1,z_{\nu})G_0(\boldsymbol{q}-\boldsymbol{k}_1,\omega_n-z_{\nu}) \right]\biggl\},\nonumber
\end{eqnarray}
with $\varepsilon\rightarrow +0$. These equations for the one-particle and gap functions provide the basis for the study of BCS suerconductivity. To simplify the analysis further and to derive approximate analytical results, it is assumed that the Cooperons exist only in a very narrow zero-frequency mode so that the gap function is given by
\begin{equation}
\Delta(\boldsymbol{k}\boldsymbol{k}^{\prime}\boldsymbol{q}|\omega_n)=\frac{1}{2}\sum\limits_{s_1,s_2}
\Delta_{s_1s_2,s_1s_2}(\boldsymbol{k}\boldsymbol{k}^{\prime}\boldsymbol{q}|\omega_n)
=-i\hbar\beta\delta_{n,0}\delta_{\boldsymbol{q},\boldsymbol{0}}\Delta(\boldsymbol{k},\boldsymbol{k}^{\prime}).
\end{equation}
Adopting an effective expression for the Hartree-Fock GF of the form
\begin{equation}
G_0(\boldsymbol{k}_1,z_{\nu})=\frac{1}{\hbar z_{\nu}-\varepsilon(\boldsymbol{k})},
\end{equation}
with $\varepsilon(\boldsymbol{k})$ denoting the kinetic energy of carriers and restricting to a short-range attraction with the coupling constant $g$, the gap equation reduces to
\begin{equation}
1=\frac{g}{\beta}\sum\limits_{\boldsymbol{k},\nu}\frac{1}{(\hbar z_{\nu}-\mu)^2-E(\boldsymbol{k})^2},\quad
E(\boldsymbol{k})=\sqrt{(\varepsilon(\boldsymbol{k})-\mu)^2+\Delta(\boldsymbol{k},\boldsymbol{k})},
\end{equation}
which is exactly the result of the well-known BCS theory.

Recapitulating this derivation, it is tempting to speculate that in the crossing symmetric particle-hole channels III and IV there also exist specific, physically meaningful two-particle excitations such as, e.g., the paramagnon \cite{Weiner}. Both a numerical and analytical study of all these contributions to the correlated two-particle GF, which are given by
\begin{eqnarray}
&&G_c(12,1^{\prime}2^{\prime})\sim i\hbar\int{\rm d}\overline{1}{\rm d}\overline{2} V(\overline{1}-\overline{2})\\
&&\times\frac{1}{2}\biggl\{G_0(1,\overline{1})G(\overline{2},1^{\prime})G_c(\overline{1}2,\overline{2}^{+}2^{\prime})
+G_0(2,\overline{1})G(\overline{2},1^{\prime})G_c(1\overline{1},\overline{2}^{+}2^{\prime})\nonumber\\
&&+G_0(1,\overline{1})G(\overline{2},2^{\prime})G_c(\overline{1}2,1^{\prime}\overline{2}^{+})
+G_0(2,\overline{1})G(\overline{2},2^{\prime})G_c(1\overline{1},1^{\prime}\overline{2}^{+})
\biggl\}\nonumber
\end{eqnarray}
\begin{eqnarray}
&&G_c(12,1^{\prime}2^{\prime})\sim -i\hbar\int{\rm d}\overline{1}{\rm d}\overline{2} V(\overline{1}-\overline{2})\\
&&\times\frac{1}{2}\biggl\{G_0(1,\overline{1})G(\overline{1},1^{\prime})G_c(\overline{2}2,\overline{2}^{+}2^{\prime})
+G_0(1,\overline{1})G(\overline{1},2^{\prime})G_c(\overline{2}2,1^{\prime}\overline{2}^{+})\nonumber\\
&&+G_0(2,\overline{1})G(\overline{1},1^{\prime})G_c(1\overline{2},\overline{2}^{+}2^{\prime})
+G_0(2,\overline{1})G(\overline{1},2^{\prime})G_c(1\overline{2},1^{\prime}\overline{2}^{+})
\biggl\}\nonumber
\end{eqnarray}
for the particle-hole channels, would contribute to a systematic analysis of the pair approximation in the many-body theory. In particular, such an approach would be applicable to critical regions of phase transitions in the vicinity of singularities in correlation functions and would allow the identification of symmetry breaking order parameters.

\section{Summary}
The conventional parquet equations couple two-particle scattering channels, which are described by partial vertex functions of thermodynamic GFs to account for the crossing symmetry. This many-body theory is completely exact, when all irreducible vertices are included in the irreducible vertex function. Unfortunately, the full parquet equations cannot be solved without truncations. Any approximate treatment of the parquet equations, however, depends sensitively on the grouping of correlation functions in the hierarchy of equations. An attractive alternative to the conventional equation of motions for thermodynamic GFs represents the cumulant expansion, which generates a chain of equations for correlated GFs. It is an advantage of the $G_c$ hierarchy that it creates step by step self-consistent $n$-particle approximations. For instance, the pair approximation arises in a natural way, when the correlated three-particle GF is neglected. Unfortunately, this procedure has a serious defect: the resulting equations for the correlated GFs do not exhibit crossing symmetry. Therefore, an extension of basic equations for correlated GFs is needed that is in line with this fundamental symmetry. It was the aim of this paper to present a set of parquet equations that represents a crossing-invariant cumulant expansion, from which the conventional parquet equations are straightforwardly recovered. Of particular interest is the linearized version of this parquet approach for correlated GFs.


%

%

\end{document}